\documentclass[aps,twocolumn,showpacs,superscriptaddress,amsmath]{revtex4}
\usepackage{graphicx}

\begin{document}
\date{August 2002}
\title{
Shot noise and  transport 
in small quantum cavities with large openings
}
\author{R. G. Nazmitdinov}
\affiliation{Max-Planck-Institut f\"ur Physik komplexer
Systeme, D-01187 Dresden, Germany}
\affiliation{Departament de F{\'\i}sica, Universitat de les Illes Balears, E-07071
Palma de Mallorca, Spain}
\affiliation{Bogoliubov Laboratory of Theoretical Physics, Joint Institute
for Nuclear Research, 141980 Dubna, Russia}
\author{ H.-S. Sim }
\affiliation{Max-Planck-Institut f\"ur Physik komplexer
Systeme, D-01187 Dresden, Germany}
\author{H. Schomerus}
\affiliation{Max-Planck-Institut f\"ur Physik komplexer
Systeme, D-01187 Dresden, Germany}
\author{I. Rotter}
\affiliation{Max-Planck-Institut f\"ur Physik komplexer
Systeme, D-01187 Dresden, Germany}

\begin{abstract}
We present a dynamical analysis of the transport through small quantum cavities
with large openings.
The systematic suppression of shot noise is used to distinguish
direct, deterministic from indirect, indeterministic transport processes.
The analysis is based on quantum mechanical 
calculations of $S$ matrices and their poles 
for quantum billiards with convex boundaries of different shape  
and two open channels in each of the two attached leads. 
Direct processes are supported when special states couple strongly 
to the leads, and can
result in deterministic transport as signified
by a striking system-specific suppression 
of shot noise. 
\end{abstract}

\pacs{73.23.Ad, 05.45.Mt, 72.70.+m}
\maketitle

Transport through quantum devices is one of the topical subjects
in mesoscopic physics. 
Diagrammatic perturbation theory and
random-matrix theory (RMT) 
predict 
that the conductance varies from sample to sample
by a universal small amount,
if details of the quantum device and its coupling to the 
electronic reservoirs are not important \cite{diag,bljab} (for reviews see Refs.\ 
\cite{ben,alh}).
These universal conductance fluctuations have 
been observed experimentally for disordered quantum wires \cite{webb}
and quantum dots \cite{marcus}.
RMT assumes that the dwell time $\tau_{\rm dwell}$
of electrons in the system is large enough to wash out
all system-specific details, and supposes that
the conductance is determined by
a random coupling of the scattering states to the leads.
Recent studies clearly demonstrate 
that the conductance of small quantum dots
with very large openings, which violate the RMT conditions,
indeed only can be fully understood
on the basis of their specific properties \cite{Ak,zo}. 
This is the deep quantum mechanical regime, in which
the Heisenberg time
$\tau_{\rm H}=\hbar/\Delta$ becomes
a relevant time scale for the internal dynamics,
where $\Delta$ is the mean resonance spacing.
Under these conditions the coupling between specific
states of the cavity and the channel wave functions
depends
sensitively and non-universally on the
position of the attached leads \cite{1,2}.
For instance, an appropriate attachment of the leads to the 
Bunimovich stadium billiard, a paradigm for 
quantum chaotic systems, gives rise to a family of 
short-lived whispering gallery modes (WGM) 
related to the shortest classical paths connecting both leads. 
These special states have small lifetimes $\tau_{\rm dwell}\ll\tau_{\rm H}$
and correspond to broad overlapping 
resonances, which influence the transport over a large 
range of energies.
As a consequence, the conductance of the quantum cavity significantly exceeds 
the prediction of RMT  \cite{2}.

It is well established that shot noise
(the zero-frequency current-current
correlations caused by the discreteness of electric charge)
provides valuable
complementary information not contained in the conductance
(for a review see Ref.\ \cite{3}). 
It was found that correlations due to Fermi statistics 
suppress the shot-noise power $P$ by a
factor ${\cal F}=P/P_0$
below the maximal value $P_0=2e G_0 V$ of incoherent transport, 
with $e$ the unit of charge, $G_0$ the series
conductance of the
two point contacts, and $V$ the applied voltage.
The universal predictions for the shot-noise suppression factor
are ${\cal F}=1/3$ for  noninteracting electrons in diffusive wires \cite{4,44} 
and ${\cal F}=1/4$ for quantum dots \cite{5,55}, respectively.  

Classically, the zero-temperature shot noise vanishes
because of the deterministic nature of classical transport.
Based on the semiclassical approach to quantum 
transport in classically chaotic systems,
it was predicted that the shot noise indeed can be reduced
below ${\cal F}=1/4$
if the Ehrenfest time $\tau_E$ (the mean time for
wave packets to dissolve)
is larger than $\tau_{\rm dwell}$
\cite{Ag}.
Recently, this prediction has been verified experimentally 
by using chaotic cavities,
where the time that electrons dwell inside can be tuned \cite{Ob}.
The analogue suppression in soft chaotic systems
has been considered in Ref.\ \cite{7}.

In the present paper we investigate the shot noise in
the deep quantum mechanical regime,
which is opposite to the semiclassical limit 
of Refs.\ \cite{Ag,Ob,7}.
We find that the bands of broad resonances
(formed by appropriate attachment of the leads)
 support direct, deterministic
transport channels with dwell times $\tau_{\rm dwell}$ less than the
wave-packet dispersion time $\tau_{E}$.
This is signified by a striking
suppression of the 
shot noise,  even when the corresponding 
classical dynamics in the quantum cavity is chaotic.
On the other hand,  resonances with large dwell times
are indeterministic and contribute to the shot noise as predicted
by RMT.

\begin{figure*}
\includegraphics[width=0.7\textwidth]{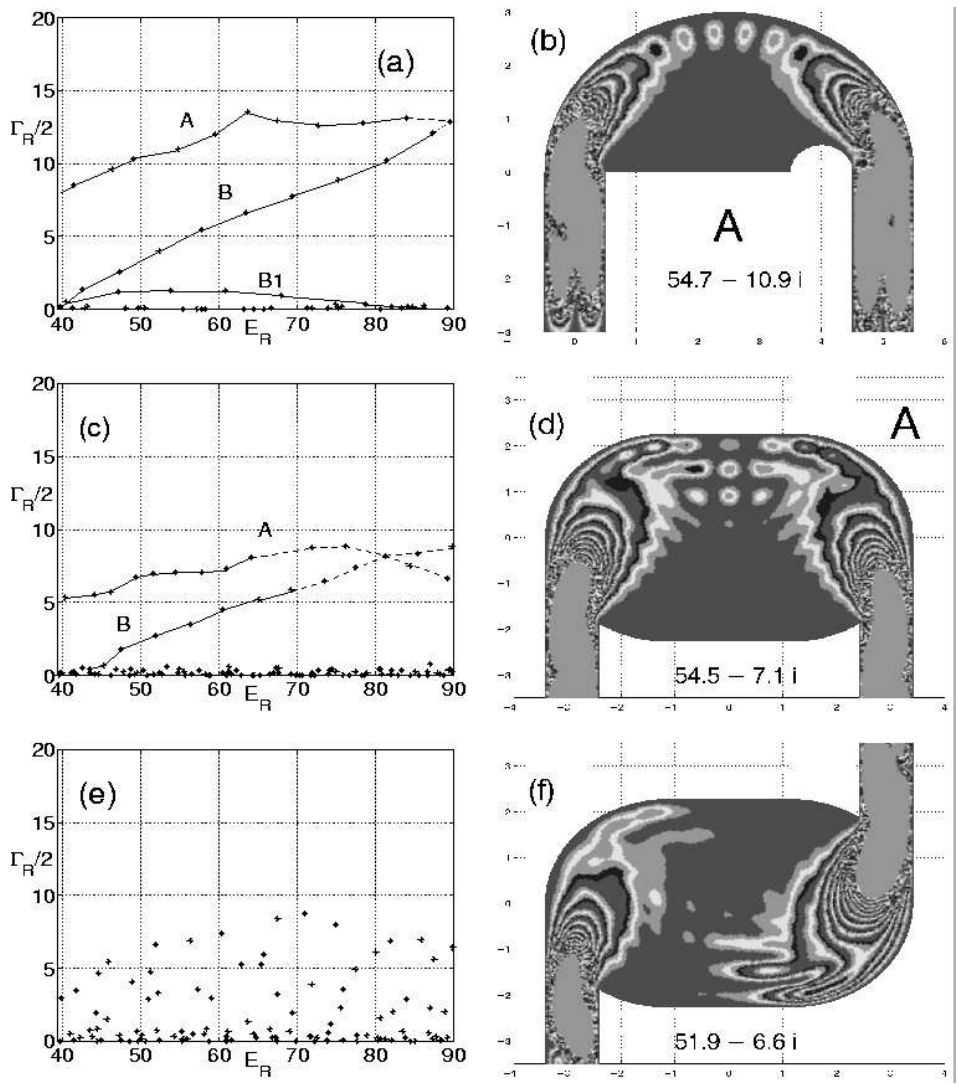}
\caption{
The poles ${\cal E}_R=E_R-i\Gamma_R/2$ of the $S$ matrix (left panels) and representative pictures
$|\Phi_R|^2$ of the wave functions of 
short-lived states (right panels)
for the billiards SIS (a, b), Bun1 (c, d), and  Bun2 (e, f).
Poles of the $S$ matrix (denoted by stars) far from the real axis
are  connected by lines for guiding the eye.
The resonance states are shown for the energy interval 
$4\pi^2\leq E \leq 9\pi^2$,  between 
the second and the third threshold for propagating modes in the leads.
The energies and widths are in units 
$\hbar^2/(2md^2)=1$, where $d=1$ is the width 
of the leads.
}
\label{fig1}
\end{figure*}

In order to study the
role of special states for the transport in small quantum systems
with large openings,
we consider cavities with convex boundaries and two
attached leads, where each lead supports $N=2$ propagating modes. 
We analyze
two stadium billiards of Bunimovich
type [length of the linear segment $L=3\pi/(\pi+1)$ and radius $R=L$]
with different positions of the leads [Bun1 and Bun2, see Fig.\
\ref{fig1}(d,f)], and also
consider a semicircle $(R=3)$ with an 
internal scatterer
 [SIS, see Fig.\
\ref{fig1}(b)]. The Bunimovich stadium is a prototype of chaotic
motion, while the SIS is most suitable 
for the propagation of the WGM \cite{1}. 

The poles of the $S$ matrix are obtained by the method of
exterior complex scaling \cite{Mois} in combination with
the finite element method (for details see Ref.\ \cite{Per}). 
The $S$ matrix itself is calculated in small energy steps by
directly solving the
Schr\"odinger equation in a discretized space, according to the method
suggested in Ref.\ \cite{Ando}.
The poles of the $S$ matrix obtained by the numerical calculation are shown
in the left panels of Fig.\ \ref{fig1}.
The plots demonstrate the formation of bands of overlapping resonances
denoted by A and B for the SIS and the Bun1 billiard,
while no evident band is formed in the Bun2 billiard.
Another band B1 of resonances can be identified for the SIS, but these
resonances do not overlap.

The bands of overlapping resonances
in the SIS and in the Bun1 billiard are the result of
resonance trapping \cite{Rot}, a phenomenon in which few states
accumulate the major part of the total sum of the widths 
$\sum_R \Gamma_R(E)$ (which is fixed by  a sum rule;
for quantum billiards see Refs.\ \cite{1,2}).
The states from band A and B are related to the first and second
transverse excitation of the
propagating modes in the leads, respectively.
The resonance wave function
$|\Phi_R(x,y)|^2$ displays a strong localization along the
convex boundary, which is
a characteristic feature of the
WGM \cite{2}.
In the SIS, these trajectories almost exclusively correspond to one bounce
at the convex boundary, while for the Bun1 billiard, 
there is also a small contribution of trajectories with two bounces.
In the SIS the WGM accumulate 
$R=\sum \Gamma_{WGM}/\sum_i \Gamma_i > 98\%$ of the total sum 
of widths $\sum \Gamma_i$ of all states, while
in the Bun1 billiard they
accumulate
a fraction $R=93\%$ of the total sum.
For these direct processes one can suppose that
the Ehrenfest time $\tau_E$ is larger than the dwell time  
$\tau_{\rm dwell}$, and this will be demonstrated, indeed, by our dynamical
analysis.

The bands of WGM are inhibited in the
Bun2 billiard [Fig.\ \ref{fig1}(e,f)] by destructive interference,
since the coupling matrix elements of the WGM 
with the channel wave functions 
have different phases for different leads.
As we will demonstrate, the resonances in the Bun2
comply with an Ehrenfest time $\tau_E<\tau_{\rm dwell}$,
corresponding to indeterministic processes.
This is also the case for the
long-living resonances  in the SIS and Bun1 billiards.

\begin{figure}
\includegraphics[angle=-90,width=0.85\columnwidth]{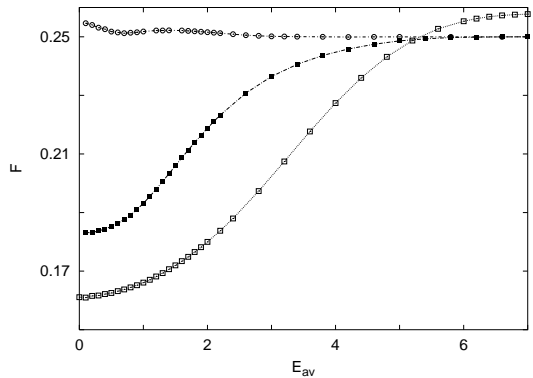}
\caption{
Shot-noise suppression factor $\cal{F}$
versus the value of the energy-averaging interval $E_{\rm av}$
for the three different billiards
SIS (open square), Bun1 (filled square), and Bun2 (open circle).
}
\label{fig2}
\end{figure}

The suppression factor 
${\cal{F}}=(2/N) {\rm tr}[T(1-T)]$
is determined by the eigenvalues (transmission probabilities) of the
matrix $T= t^\dagger t$, where $t$ is the transmission matrix
\cite{6,44}.
The closer the transmission probabilities are to the classical values
$0$ or $1$ of completely deterministic dynamics, 
the smaller is the shot
noise. 
To get a thorough insight into the interplay of deterministic and
indeterministic dynamics, we employ the
procedure recently developed in Ref.\ \cite{7}.
The major element of this procedure is the purposeful replacement of
system-specific dynamics with dwell times
$\tau_{\rm dwell}\gtrsim \tau_{\rm av}$
by indeterministic processes.
In the first step, the system-specific details are eliminated by
averaging the
scattering matrix $S$ over an energy range 
$[E_0-E_{\rm av}/2,E_0+E_{\rm av}/2]$ of width $E_{\rm av}=\hbar/\tau_{\rm
av}$,
\begin{equation}
\bar{S}(E_{\rm av};E_0)=
\frac{1}{E_{\rm av}}\int_{E_0-E_{\rm av}/2}^{E_0+E_{\rm av}/2}S(E)dE.
\end{equation}
Indeterministic processes then are introduced by coupling the system
to an
auxiliary indeterministic system, with
a scattering matrix $S_0$
taken from the appropriate circular ensemble of RMT.
The composed system
with scattering matrix 
\begin{equation}
S^{\prime}(E_{\rm av};E_0;S_0)= \bar{S}(E_{\rm av};E_0)+
{\cal T}^{\prime}(1-S_0 {\cal{R}})^{-1}S_0{\cal T}
\end{equation}
is a member of the
Poisson kernel of RMT \cite{Pos,bro,bar}.
The matrix $S^{\prime}$ must be unitary, which
determines the coupling matrices ${\cal T}$, 
${\cal T}^{\prime}$ and ${\cal{R}}$.
We numerically calculate the mean suppression factor ${\cal{F}}(E_{\rm av})$ 
for fixed values of $E_{\rm av}$ by first averaging over
the random matrix $S_0$ within each Poisson kernel, and then averaging
over $E_0$ within a range $(E_0^{\rm min}, E_0^{\rm max})$.

\begin{figure}
\includegraphics[angle=-90,width=0.85\columnwidth]{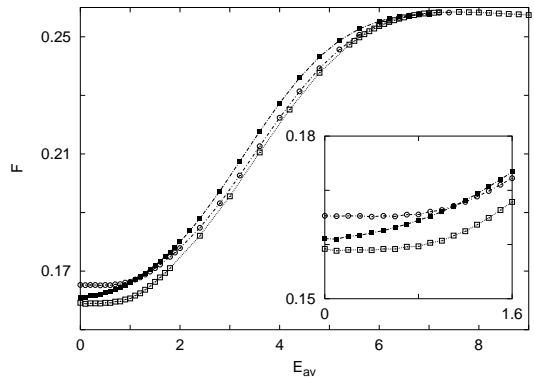}
\caption{
Shot-noise suppression factor ${\cal{F}}$ of the SIS billiard
as a function
of the energy-averaging interval $E_{\rm av}$.
Different curves correspond to the different lower energy bounds
$E_0^{\rm min}=40$ (filled square),
$E_0^{\rm min}=52.5$ (open circle), and
$E_0^{\rm min}=65$ (open square).
}
\label{fig3}
\end{figure}

The result for the suppression factor ${\cal F}(E_{\rm
av})$ in the three billiards is shown in Fig.\ \ref{fig2}.
Here $E_0^{\rm min}=40$ and $E_0^{\rm max}=90$  are fixed to the values where
the second and the third propagating mode opens in each lead, respectively.
The value ${\cal{F}}(E_{\rm av}=0)$ is the shot-noise suppression factor
of the billiards themselves, without
any modification of their dynamics.
The shot noise in the  billiard Bun2 corresponds well to the
universal prediction ${\cal F}=1/4$ of RMT, 
from which we conclude that $\tau_{\rm dwell}\gtrsim\tau_{E}$
and that the dynamics is indeterministic. 
For the other two billiards, 
the suppression factor is significantly reduced below the universal
value.
We now argue that this can be attributed to the
WGM, which support deterministic transport.

The modes that lead to shot-noise reduction can be identified from the
dependence of ${\cal{F}}(E_{\rm av})$ 
on $E_{\rm av}$, which indicates the sensitivity to replacement of
system-specific processes by indeterministic ones.
For the  billiards SIS and Bun1,
the function $\cal{F}$ increases 
monotonically 
with increasing 
contribution of the random dynamics (increasing $E_{\rm av}$)
and eventually
approaches the universal prediction of RMT.
However, the
slope of ${\cal{F}}(E_{\rm av})$ for small $E_{\rm av}$
almost vanishes. Hence, without any consequences for the shot noise
the long-living resonances
(trapped by the WGM) can be replaced by (and hence are equivalent to)
indeterministic processes.
On the other hand, the suppression factor rises significantly when the
energy averaging window $E_{\rm av}$ becomes of the order of the width
$\Gamma_R$ of the WGM modes in each of the two billiards (cf.\ Fig.\ \ref{fig1}),
which hence support deterministic transport.

The position of the attachment of the leads plays a decisive role, since it
determines the character of the system-specific broad resonances.
In the Bun2 billiard, the suppression factor $\cal{F}$ is
almost independent of $E_{\rm av}$ and hence
completely insensitive 
to the replacement
of system-specific properties
by indeterministic dynamics.
Even the short-living resonances no longer
correspond to deterministic classical transport.
The remarkable difference in the behavior 
of the suppression factor $\cal{F}$ for the Bun1 and Bun2 billiards
arises from the
different position of the leads attached to the 
same quantum system, which results in the formation of direct,
deterministic transport channels in the first case, while no such
channels are formed in the second case.

Both in the SIS and in the Bun1 billiard, two bands A, B of broad resonances have been identified.
We now analyze these bands in more detail for the SIS.
Figure \ref{fig3} displays the result for the suppression factor
for three different values of $E_0^{\rm min}$, while $E_0^{\rm
max}=90$ remains fixed to the value where the third propagating mode
opens in the leads. In this way we focus on the contribution of the WGM
from different
parts of the bands.
The lowest and the highest suppression factor ${\cal{F}}$ 
are obtained if we only consider WGM
with energies $E_R\geq E_0^{\rm min}=65$
and $E_R\geq E_0^{\rm min}=53$, respectively 
[see Fig.\ \ref{fig1}(a) and the inset in Fig.\ \ref{fig3}]. 
The difference between the largest and the smallest values is 
small, since the dynamics is mainly
determined by the states of the band A, which are strongly localized 
along the convex boundary and all have similar width.
The widths of the states from the family B
depend almost linearly on their energy,
and become small when one approaches the threshold $E_R=40$ of the second
propagating mode in the leads.
For $E_0^{\rm min}=40$
the resonances of type B with smallest width are
included into the averaging interval, and
the slope of the suppression factor ${\cal F}(E_{\rm av})$
for very small values of $E_{\rm av}$ increases noticeably.
This shows that even the longest-living resonances in
family B support system-specific direct, deterministic processes.

Summarizing, we analyzed the transport through small quantum cavities with
large openings, deep in the quantum regime.
We observed an essential suppression of the shot noise
when the transport is dominated
by system-specific broad resonances 
that support direct processes
well described by deterministic classical dynamics.
These deterministic transport channels can exist
even if the closed system manifests typical properties of
chaotic dynamics, and 
their formation sensitively depends on the precise position of the leads.

We gratefully acknowledge discussions with Konstantin Pichugin.

\end{document}